\documentclass{article}

\usepackage{graphicx}
\usepackage{color}
\usepackage[left=3cm,right=3cm,top=3cm,bottom=2cm]{geometry}

\title{On the Charge to Mass Ratio of Neutron Cores and Heavy Nuclei}

\author{B. Patricelli$^{1,2}$, M. Rotondo$^{1,2}$, R. Ruffini$^{1,2,3}$ \\\\
$^{1}$ ICRAnet and ICRA, Ple. della Repubblica 10, Pescara, Italy
\\$^{2}$ University of Rome ``La Sapienza'', Piazzale Aldo Moro 5, Roma, Italy 
\\ $^{3}$ ICRAnet, University of Nice, 28 avenue de Valrose, Nice, France
}

\date{\today}

\begin{document}

\maketitle

\begin{abstract}
We determine theoretically the relation between the total number of
protons $N_{p}$ and the mass number $A$ (the charge to mass  ratio) of
nuclei and neutron cores with the model recently proposed by Ruffini
et al. (2007) and we compare it with other $N_p$ versus $A$ relations: the empirical
one, related to the Periodic Table, and the semi-empirical relation,
obtained by minimizing the Weizs\"{a}cker mass
formula. We find that there is a very good agreement between all the
relations for values of $A$ typical of nuclei, with differences of the
order of per cent. Our relation and the semi-empirical one are in
agreement up to $A\sim 10^4$; for higher values, we find that the
two relations differ. We interprete the different behaviour of our
theoretical relation  as a result of the penetration of electrons
(initially confined in an external shell) inside the core, that
becomes more and more important by increasing $A$; these effects are
not taken into account in the semi-empirical mass-formula.
\end{abstract}

\maketitle

\section{Introduction}

It is well known that stable nuclei are located, in the $N_n$-$N_p$ plane (where $N_n$ and $N_p$ are the total number of neutrons and protons respectively), in a region that, for small values of $N_p$, is almost a line well described by the relation $N_n=N_p$.\\
In the past, several efforts have been made to explain theoretically this property, for example with the liquid drop model of atoms, that is based on two properties common to all nuclei: their mass densities and their binding energies for nucleons are almost indipendent from the mass number $A=N_n+N_p$ \cite{segre}. This model takes into account the strong nuclear force and the Coulombian repulsion between protons and explains different properties of nuclei, for example the relation between $N_p$ and $A$ (the charge to mass ratio). \\
In this work we derive theoretically the charge to mass ratio of nuclei and extend it to neutron cores  (characterized by higher values of $A$) with the model of Ruffini et al. \cite{Elec2007}, \cite{ArtM}.
We consider systems composed of degenerate neutrons, protons and electrons and we use the relativistic Thomas-Fermi equation and the equation of $\beta$-equilibrium to determine the number density and the total number of these particles, from which we obtain the relation between $N_p$ and $A$.

\section{The theoretical model}
Following the work of Ruffini et al. \cite{Elec2007}, \cite{ArtM}, we describe nuclei and neutron cores as spherically symmetric systems composed of degenerate protons, electrons and neutrons and impose the condition of  global charge neutrality.\\
We assume that the proton's number density $n_p(r)$ is constant inside the core ($r\leq R_C$) and vanishes outside the core ($r>R_C$):
\begin{equation}\label{eq:denp}
n_p(r)=\left( \frac{3 N_p}{4 \pi R_C^3}\right) \theta(R_C-r),
\end{equation}
where $N_p$ is the total number of protons and $R_C$ is the core-radius, parametrized as
\begin{equation} \label{eq:RC}
R_C=\Delta \frac{\hbar}{m_\pi c} N_p^{1/3}.
\end{equation}
We choose $\Delta$ in order to have $\rho \sim \rho_N$, where $\rho$ and $\rho_N$ are the mass density of the system and the nuclear density respectively ($\rho_N=2.314\cdot 10^{14} g/cm^{-3}$).\\
The electron number density $n_e(r)$ is given by
\begin{equation}
n_e(r)=\frac{1}{3 \pi^2 \hbar^3}\left[ p_e ^F(r)\right]^3,
\end{equation}
where $p_e ^F(r)$ is the electron Fermi momentum. It can be calculated from the condition of equilibrium of Fermi degenerate electrons, that implies the null value of their Fermi energy $\epsilon_e ^F(r)$
\begin{equation} \label{eq:efe}
\epsilon_e ^F(r)=\sqrt{[p_e ^F(r) c]^2+m_e^2 c^4}-m_e c^2+V_c(r)=0,
\end{equation}
where $V_c(r)$ is the Coulomb potential energy of electrons.

From this condition we obtain
\begin{equation}
p_e ^F(r)=\frac{1}{c}\sqrt{V_c^2(r)-2m_ec^2 V_c(r)},
\end{equation}
hence the electron number density is
\begin{equation}\label{eq:dene}
n_e(r)=\frac{1}{3 \pi^2 \hbar^3 c^3}\left[V_c^2(r)-2m_ec^2 V_c(r)\right]^{3/2}.
\end{equation}
The Coulomb potential energy of electrons, necessary to derive $n_e(r)$, can be determined as follows.
Based on the Gauss law, $V_c(r)$ obeys the following Poisson equation:
\begin{equation}\label{eq:poisson}
\nabla^2 V_c(r)=-4 \pi e^2[n_e(r)-n_p(r)],
\end{equation}
with the boundary conditions $V_c(\infty)=0$, $V_c(0)=finite$. Introducing the dimensionless function $\chi(r)$, defined by the relation
\begin{equation}
V_c(r)=-\hbar c \frac{\chi(r)}{r},
\end{equation}
and the new variable $x=r b^{-1}=r \left(\frac{\hbar} {m_\pi c}\right)^{-1}$, from eq. (\ref{eq:poisson}) we obtain the relativistic Thomas-Fermi equation
\begin{eqnarray} \label{eq:RTF}
\frac{1}{3x}\frac{d^2\chi(x)}{dx^2}=-\alpha\left\{\frac{1}{\Delta^3} \theta(x_c-x)-\frac{4}{9 \pi}\left[\frac{\chi^2(x)}{x^2}+2 \frac{m_e}{m_{\pi}} \frac{\chi(x)}{x}\right]^{3/2} \right\}.
\end{eqnarray}
The boundary conditions for the function $\chi(x)$ are
\begin{equation}
\chi(0)=0, \qquad \chi(\infty)=0,
\end{equation}
as well as the continuity of $\chi(x)$ and its first derivative $\chi^{'}(x)$ at the boundary of the core.\\
The number density of neutrons $n_n(r)$ is
\begin{equation} \label{eq:denn}
n_n(r)=\frac{1}{3 \pi^2 \hbar^3}\left[ p_n ^F(r)\right]^3,
\end{equation}
where $p_n ^F(r)$ is the neutron Fermi momentum. It can be calculated with the condition of equilibrium between the processes
\begin{equation}
e^-+p\rightarrow n+\nu_e;
\end{equation}
\begin{equation}
n \rightarrow p+e^-+\bar{\nu_e},
\end{equation}
Assuming that neutrinos escape from the core as soon as they are produced, this condition (condition of $\beta$-equilibrium) is
\begin{equation} \label{eq:betaeq}
\epsilon ^F_e(r)+\epsilon ^F_p(r)=\epsilon ^F_n(r).
\end{equation}
Eq. (\ref{eq:betaeq}) can be explicitly written as
\begin{eqnarray} \label{eq:betaeq2}
\sqrt{[p_p ^F(r) c]^2+m_p^2 c^4}-m_p c^2-V_c(r)=\sqrt{[p_n ^F(r) c]^2+m_n^2 c^4}-m_n c^2.
\end{eqnarray}

\section{$N_p$ versus $A$ relation}
Using the previous equations, we derive $n_e(r)$, $n_n(r)$ and $n_p(r)$ and, by integrating these, we obtain $N_e$, $N_n$ and $N_p$. We also derive a theoretical relation between $N_p$ and $A$ and
we compare it with the data of the Periodic Table and with the semi-empirical relation
\begin{equation}\label{eq:semiemp}
N_p=\left(\frac{A}{2}\right)\cdot \frac{1}{1+\left(\frac{3}{400}\right) \cdot A^{2/3}}
\end{equation}
that, in the limit of low $A$, gives the well known relation $N_p=A/2$ \cite{segre}.\\
Eq. (\ref{eq:semiemp}) can be obtained by minimizing the semi-empirical mass formula, that was first formulated by Weizs\"{a}cker in 1935 and is based on empirical measurements and on theory (the liquid drop model of atoms). \\
The liquid drop model approximates the nucleus as a sphere composed of protons and neutrons (and not electrons) and takes into account the Coulombian repulsion between protons and the strong nuclear force. Another important characteristic of this model is that it is based on the property that the mass densities of nuclei are approximately the same, indipendently from $A$ \cite{eisberg}. In fact, from scattering experiments it was found the following expression  for the nuclear radius $R_N$:
\begin{equation} \label{eq:rn}
R_N=r_0 A^{1/3},
\end{equation}
with $r_0=1.2$ fm. Using eq. (\ref{eq:rn}) the nuclear density can be write as follows:
\begin{equation}\label{eq:dennucl}
\rho_N=\frac{A m_N}{V}=\frac{3 A m_N}{4 \pi r_0^3 A}=\frac{3 m_N}{4 \pi r_0^3},
\end{equation}
where $m_N$ is the nucleon mass.
From eq. (\ref{eq:dennucl}) it is clear that nuclear density is indipendent from $A$, so it is constant for all nuclei. \\
The property of constant density for all nuclei is a common point with our model: in fact, we choose $\Delta$ in order to have the same mass density for every value of $A$; in particular we consider the case $\rho \sim \rho_N$, as previously said.

\begin{figure}
	\begin{center}
        \includegraphics[scale=0.5]{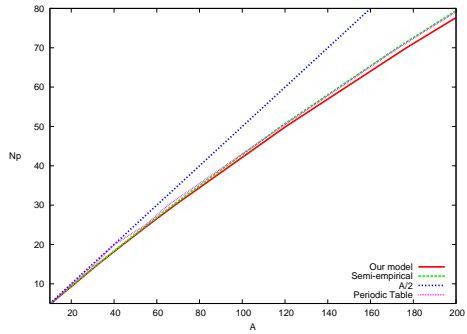}
    \caption{\footnotesize{The $N_p-A$ relation obtained with our model and with the semi-empirical mass formula, the $N_p=A/2$ relation and the data of the Periodic Table; relations are plotted for values of $A$ from 0 to 200.}}\label{fig:PT}
   \end{center}
\end{figure}

\begin{figure}
	\begin{center}
        \includegraphics[scale=0.5]{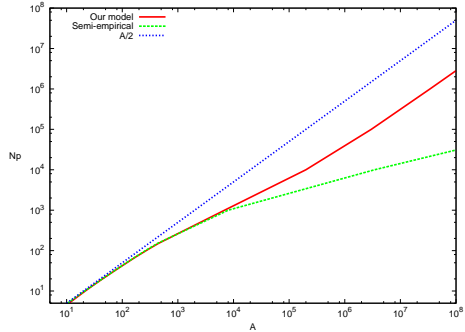}
    \caption{\footnotesize{The $N_p-A$ relation obtained with our model and with the semi-empirical mass formula and the N$_p=A/2$ relation; relations are plotted for values of $A$ from 0 to 10$^8$. It is clear how the semi-empirical relation and the one obtained with our model are in good agreement up to values of $A$ of the order of 10$^4$; for greater values of $A$ the two relation differ because our model takes into account the penetration of electrons inside the core, which is not considered in the semi-empirical mass formula.}}\label{fig:PE}
  \end{center}
\end{figure}

\begin{figure}
	\begin{center}
        \includegraphics[scale=0.5]{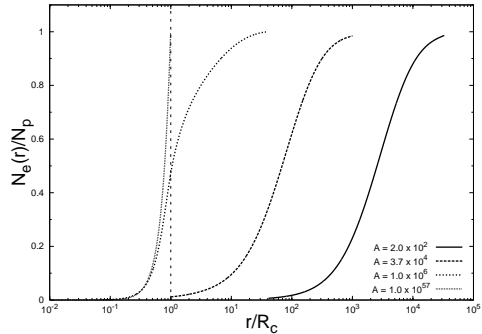}
    \caption{\footnotesize{The electron number in units of the total proton number $N_p$ as function of the radial distance in units of the core radius R$_C$, for different values of $A$. It is clear that, by increasing the value of A, the penetration of electrons inside the core increases. Figure from Ruffini et al.  \cite{Elec2007}.}}\label{fig:elec07}
  \end{center}
\end{figure}

In table (\ref{tab:all}) are listed some values of $A$ obtained with our model and the semi-empirical mass formula, as well as the data of the Periodic Table; in fig. (\ref{fig:PT}) and (\ref{fig:PE}) it is shown the comparison between the various $N_p-A$ relations.

It is clear that there is a good agreement between all the relations for values of $A$ typical of nuclei, with differences of the order of per cent. Our relation and the semi-empirical one are in agreement up to $A\sim 10^4$; for higher values, we find that the two relations differ. We interprete these differences as due to the effects of penetration of electrons inside the core [see fig. (\ref{fig:elec07})]: in our model we consider a system composed of degenerate protons, neutrons and electrons. For the smallest values of $A$, all the electrons are in a shell outside the core; by increasing $A$, they progressively penetrate into the core \cite{Elec2007}. These effects, which need the relativistic approach introduced in \cite{Elec2007}, are not taken into account in the semi-empirical mass formula.

\begin{figure}
	\begin{center}
        \includegraphics[scale=0.5]{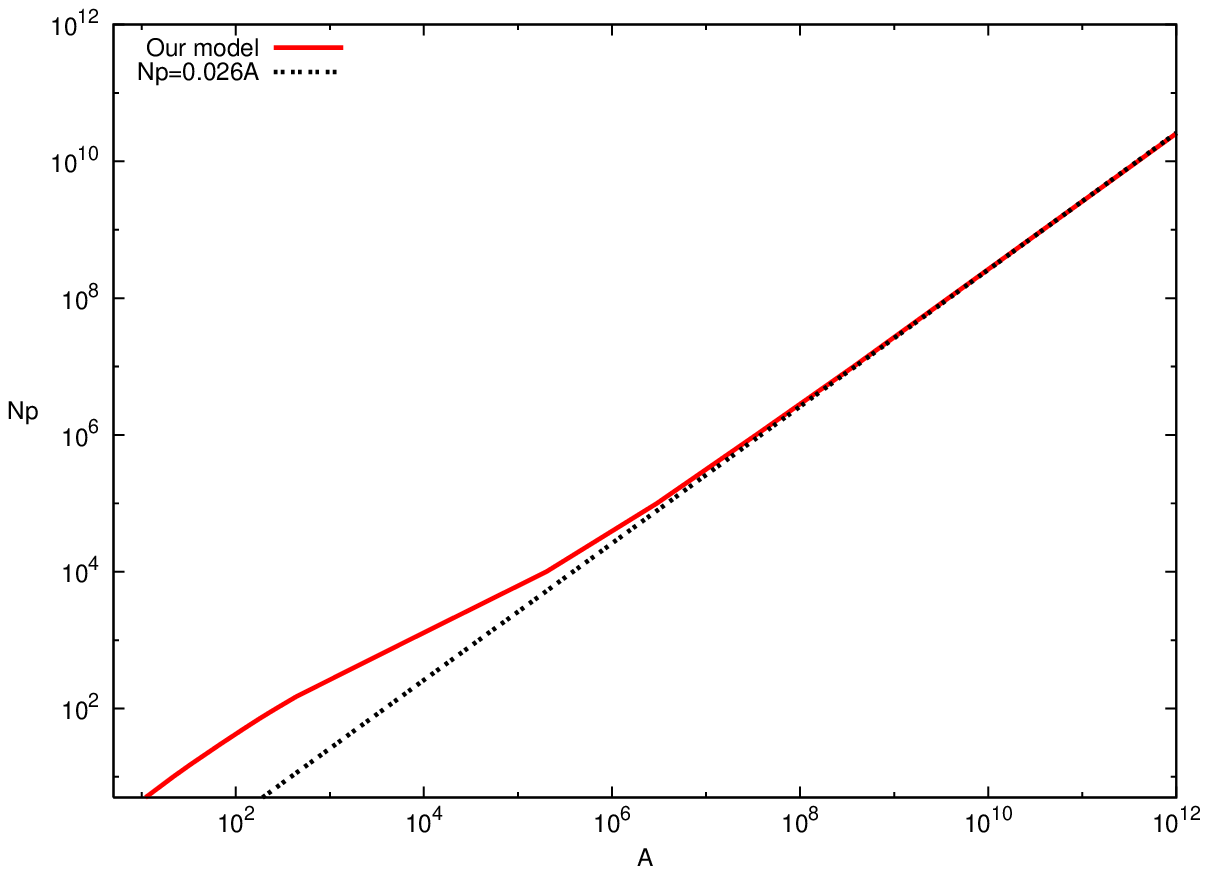}
    \caption{\footnotesize{The $N_p-A$ relation obtained with our model and the asymptotic limit $N_p=0.026 A$}}\label{fig:asym}
   \end{center}
\end{figure}

\begin{table}\label{tab:all}
\begin{center}
        \begin{tabular}{|cccc|}
        \hline
        \multicolumn{1}{|c|}{\bf{$N_p$}}& \multicolumn{1}{|c|} {\bf{A$_M$}}& \multicolumn{1}{|c|} {\bf{A$_{PT}$}}& \multicolumn{1}{|c|} {\bf{A$_{SE}$}}\\
        \hline
        \multicolumn{1}{|c|}{5}& \multicolumn{1}{|c|} {10.40}& \multicolumn{1}{|c|} {10.811}& \multicolumn{1}{|c|} {10.36}\\
        \hline
        \multicolumn{1}{|c|}{10}& \multicolumn{1}{|c|} {21.59}& \multicolumn{1}{|c|} {20.183}& \multicolumn{1}{|c|} {21.15}\\
        \hline
        \multicolumn{1}{|c|}{15}& \multicolumn{1}{|c|} {32.58}& \multicolumn{1}{|c|} {30.9738}& \multicolumn{1}{|c|} {32.28}\\
        \hline
        \multicolumn{1}{|c|}{20}& \multicolumn{1}{|c|} {44.24}& \multicolumn{1}{|c|} {40.08}& \multicolumn{1}{|c|} {43.72}\\
        \hline
        \multicolumn{1}{|c|}{25}& \multicolumn{1}{|c|} {56.17}& \multicolumn{1}{|c|} {54.938}& \multicolumn{1}{|c|} {55.45}\\
        \hline
        \multicolumn{1}{|c|}{30}& \multicolumn{1}{|c|} {68.43}& \multicolumn{1}{|c|} {65.37}& \multicolumn{1}{|c|} {67.46}\\
        \hline
        \multicolumn{1}{|c|}{50}& \multicolumn{1}{|c|} {120.40}& \multicolumn{1}{|c|} {118.69}& \multicolumn{1}{|c|} {118.05}\\
        \hline
        \multicolumn{1}{|c|}{70}& \multicolumn{1}{|c|} {176.78}& \multicolumn{1}{|c|} {173.04}& \multicolumn{1}{|c|} {172.54}\\
        \hline
        \multicolumn{1}{|c|}{90}& \multicolumn{1}{|c|} {237.41}& \multicolumn{1}{|c|} {232.038}& \multicolumn{1}{|c|} {230.79}\\
        \hline
        \multicolumn{1}{|c|}{110}& \multicolumn{1}{|c|} {302.18}& \multicolumn{1}{|c|} {271}& \multicolumn{1}{|c|} {292.75}\\
        \hline
        \multicolumn{1}{|c|}{150}& \multicolumn{1}{|c|} {443.98}& \multicolumn{1}{|c|} {}& \multicolumn{1}{|c|} {427.73}\\
        \hline
        \multicolumn{1}{|c|}{200}& \multicolumn{1}{|c|} {644.03}& \multicolumn{1}{|c|} {}& \multicolumn{1}{|c|} {617.56}\\
        \hline
        \multicolumn{1}{|c|}{250}& \multicolumn{1}{|c|} {869.32}& \multicolumn{1}{|c|} {}& \multicolumn{1}{|c|} {831.63}\\
        \hline
        \multicolumn{1}{|c|}{300}& \multicolumn{1}{|c|} {1119.71}& \multicolumn{1}{|c|} {}& \multicolumn{1}{|c|} {1071.08}\\
        \hline
        \multicolumn{1}{|c|}{350}& \multicolumn{1}{|c|} {1395.12}& \multicolumn{1}{|c|} {}& \multicolumn{1}{|c|} {1337.23}\\
        \hline
        \multicolumn{1}{|c|}{450}& \multicolumn{1}{|c|} {2019.48}& \multicolumn{1}{|c|} {}& \multicolumn{1}{|c|} {1955.57}\\
        \hline
        \multicolumn{1}{|c|}{500}& \multicolumn{1}{|c|} {2367.77}& \multicolumn{1}{|c|} {}& \multicolumn{1}{|c|} {2310.96}\\
        \hline
        \multicolumn{1}{|c|}{550}& \multicolumn{1}{|c|} {2739.60}& \multicolumn{1}{|c|} {}& \multicolumn{1}{|c|} {2699.45}\\
        \hline
        \multicolumn{1}{|c|}{600}& \multicolumn{1}{|c|} {3134.28}& \multicolumn{1}{|c|} {}& \multicolumn{1}{|c|} {3122.83}\\
        \hline
        \multicolumn{1}{|c|}{10$^3$}& \multicolumn{1}{|c|} {6.9$\cdot$10$^3$}& \multicolumn{1}{|c|} {}& \multicolumn{1}{|c|} {8$\cdot$10$^3$}\\
        \hline
        \multicolumn{1}{|c|}{10$^4$}& \multicolumn{1}{|c|} {2.0$\cdot$10$^5$}& \multicolumn{1}{|c|} {}& \multicolumn{1}{|c|} {3.45$\cdot$10$^6$}\\
        \hline
        \multicolumn{1}{|c|}{10$^5$}& \multicolumn{1}{|c|} {3.0$\cdot$10$^6$}& \multicolumn{1}{|c|} {}& \multicolumn{1}{|c|} {3.38$\cdot$10$^9$}\\
        \hline
        \multicolumn{1}{|c|}{10$^6$}& \multicolumn{1}{|c|} {3.4$\cdot$10$^7$}& \multicolumn{1}{|c|} {}& \multicolumn{1}{|c|} {3.37$\cdot$10$^{12}$}\\
        \hline
        \multicolumn{1}{|c|}{10$^7$}& \multicolumn{1}{|c|} {3.7$\cdot$10$^8$}& \multicolumn{1}{|c|} {}& \multicolumn{1}{|c|} {3.37$\cdot$10$^{15}$}\\
        \hline
        \multicolumn{1}{|c|}{10$^{10}$}& \multicolumn{1}{|c|} {3.9$\cdot$10$^{11}$}& \multicolumn{1}{|c|} {}& \multicolumn{1}{|c|} {3.37$\cdot$10$^{24}$}\\
        \hline
        \end{tabular}\caption{Different values of $N_p$ (column 1) and corresponding values of $A$ from our model  ($A_M$, column 2), the Periodic Table  ($A_{PT}$, column 3) and the semi-empirical mass formula ($A_{SE}$, column 4).}
\end{center}
\end{table}

We also note that the charge to mass ratio become constant for $A$ greater that $10^7$; in particular, it is well approximated by the relation $N_p=0.026 A$  [see fig. (\ref{fig:asym})].

\section{Conclusions}
In this work we have derived theoretically a relation between the total number of protons $N_p$ and the mass number $A$ for nuclei and neutron cores with the model recently proposed by Ruffini et al. \cite{Elec2007}, \cite{ArtM}). \\
We have considered spherically symmetric systems composed of degenerate electrons, protons and neutrons having global charge neutrality and the same mass densities ($\rho \sim \rho_N$). By integrating the relativistic Thomas-Fermi equation and using the equation of $\beta$-equilibrium, we have determined the total number of protons, electrons and neutrons in the system and hence a theoretical relation between $N_p$ and $A$.\\
We have compared this relation with the empirical data of the Periodic Table and with the semi-empirical relation,  obtained by minimizing the Weizs\"{a}cker mass formula by considering systems with the same mass densities.
We have shown that there's a good agreement between all the relations for values of $A$ typical of nuclei, with differences of the order of per cent. Our relation and the semi-empirical one are in agreement up to $A\sim 10^4$; for higher values, we find that the two relations differ. We interprete the different behaviour of our theoretical relation  as a result of the penetration of electrons (initially confined in an external shell) inside the core [see fig.(\ref{fig:elec07})], that becomes more and more important by increasing $A$; these effects, which need the relativistic approach introduced in \cite{Elec2007}, are not taken into account in the semi-empirical mass-formula.

\end{document}